\documentclass[twocolumn,secnumarabic,amsmath,amssymb,aps,nobibnotes,prl]{revtex4}

\makeatletter
\def\@dotsep{4.5}
\makeatother

\usepackage{graphicx}
\usepackage{dcolumn}
\usepackage{bm}

\usepackage{ulem}

\begin{document}


\title{
Superconductivity and Pseudogap in Quasi-Two-Dimensional Metals \\
around the Antiferromagnetic Quantum Critical Point
}%

\author{Hisashi Kondo$^{1}$ and 
T\^{o}ru Moriya$^{2}$\footnote{Professor emeritus: e-mail address: moriya-toru@topaz.ocn.ne.jp}}

\affiliation{%
$^{1}$Computational Materials Science Center, National Institute for Materials Science (NIMS), Tsukuba, Ibaraki 305-0047
\\
$^{2}$Institute for Solid State Physics, University of Tokyo, Kashiwa, Chiba, 277-8581
}%

\date{\today}%


\begin{abstract}
Spin fluctuations (SF) and SF-mediated superconductivity (SC) in quasi-two-dimensional metals around the antiferrromagnetic (AF) quantum critical point (QCP) are investigated by using the self-consistent renormalization theory for SF and the strong coupling theory for SC. We introduce a parameter $y_0$ as a measure for the distance from the AFQCP which is approximately proportional to $(x-x_{\rm c})$, $x$ being the electron (e) or hole (h) doping concentration to the half-filled band and $x_{\rm c}$ being the value at the AFQCP. We present phase diagrams in the $T$-$y_0$ plane including contour maps of the AF correlation length and AF and SC transition temperatures $T_{\rm N}$ and $T_{\rm c}$, respectively. The $T_{\rm c}$ curve is dome-shaped with a maximum at around the AFQCP. The calculated one-electron spectral density shows a pseudogap in the high-density-of-states region near $(\pi,0)$ below around a certain temperature $T^*$ and gives a contour map at the Fermi energy reminiscent of the Fermi arc. These results are discussed in comparison with e- and h-doped high-$T_{\rm c}$ cuprates. 
\end{abstract}

%
\maketitle%

%

The mechanism of high-temperature superconductivity (SC) in cuprates has been the subject of controversy for many years. In the early stage of investigations the so-called non-Fermi liquid (FL) property in the normal state, represented by $T$-linear resistivity, was a major issue in addition to the high values of the SC transition temperature $T_{\rm c}$ and its dome shape when plotted against hole (h) doping concentration \cite{rf:1}. Subsequently, the problems of the pseudo-gap in h-underdoped cuprates were brought up \cite{rf:2}. 

As a possible explanation for the non-FL properties the spin fluctuations (SF) around the two-dimensional (2D) antiferromagnetic (AF) quantum critical point (QCP) were studied by using the self-consistent theory of the coupled modes of SF or the self-consistent renormalization (SCR) theory extended to 2D systems \cite{rf:3}. This theory, which is known to give correct QC behaviors in 2D and three-dimensional (3D) systems \cite{rf:4}, explained the $T$-linear resistivity, the Curie-Weiss (CW) behavior of $1/T_1T$, $T_1$ being the nuclear spin-lattice relaxation time, and the anomalous optical conductivity. At the same time $T_{\rm c}$ for the SF-mediated $d$-wave SC was evaluated by using the spectra of SF estimated from the analysis of experimental results. The results compared well with the experimental results on optimal doping concentration ranges. Then the dependence of the value of $T_{\rm c}$ on the SF parameters and band structure parameters was studied for 2D and 3D systems \cite{rf:5, rf:6}. Fully microscopic studies of the SF mechanism were performed within the fluctuation exchange (FLEX) approximation with success in the optimal and overdoped concentration regimes \cite{rf:4}. In particular, the anomalous temperature dependence of the Hall coefficient and the neutron resonance peak in the SC phase were explained \cite{rf:7-1,rf:7-2,rf:7-3}.

As for the pseudogap phenomena, as typically observed in one-electron spectral density, many scenarios have been presented for h-underdoped systems without a consensus yet being reached \cite{rf:2}, although recent investigations on electron (e)-doped systems seem to indicate strongly that the pseudogap, observed here mainly in the AF regime above $T_{\rm N}$, is caused by the strong AF correlations of quasi-2D antiferromagnets \cite{rf:8,rf:9-1,rf:9-2,rf:10,rf:11,rf:12,rf:13}.

Under these circumstances it seems worthwhile to extend the theory of SF in quasi-2D metals around the AFQCP, in particular, to include the AF side of the QCP where the previous theory \cite{rf:3, rf:5, rf:6} has not been well extended, and discusses the above-mentioned problems. This is the purpose of the present study. 

Let us first discuss the physical properties on the AF side of the AFQCP. According to the SCR theories for purely 2D systems there is no AF order at finite temperatures even if the ground state is ordered (Mermin-Wagner theorem). The susceptibility $\chi(Q)$ obeys the CW law, $\chi(Q)=C/(T-\theta)$, with positive $\theta$ at relatively high temperatures, and below around a certain temperature $T^*$ it deviates from the CW behavior, i.e., instead of diverging toward $T = \theta$ it tends to diverge exponentially in $1/T$ toward $T=0$ \cite{rf:3}. Accordingly the AF correlation length below $T^*$ is large and diverges toward $T=0$ with decreasing temperature. The time variation of these long wavelength modes of fluctuations becomes quite slow below $T^*$. These long-ranged strong AF correlations are expected to be a possible origin of the pseudogap phenomena. In quasi-2D systems we have a finite AF critical temperature $T_{\rm N}$ owing to the weak 3D character. When the 3D character is weak and $T_{\rm N}$ is much lower than $T^*$, we have a wide temperature range of long correlation length where the pseudogap may be observed. We here present the results of explicit calculations of correlation length and $T_{\rm N}$  for various degrees of 3D character. 

We also calculate the SF-mediated $T_{\rm c}$ by using the strong coupling theory. Previous studies were concentrated on the paramagnetic side of the AFQCP \cite{rf:3, rf:5, rf:6}. Now extending calculations to the AF side we find a dome-shaped $T_{\rm c}$ curve against doping concentration. We calculate possible phase diagrams including a contour map for AF correlation length, $T^*$, $T_{\rm N}$, and $T_{\rm c}$ and compare them with experimental results for e- and h-doped cuprates. In order to see the pseudogap explicitly and to estimate $T^*$ as a pseudogap onset temperature we calculate the one-electron spectral density with varying temperature in the high density-of-states (DOS) region near $(\pi,0)$. We also calculate contour maps for the spectral density at the Fermi energy and discuss about the Fermi arc. 

The dynamical susceptibility of a quasi-2D metal around the AFQCP is calculated by using a parameterized form of the SCR theory and the following expansion form: 
\begin{eqnarray}
    1/\chi(Q+ q,\omega)=1/\chi(Q)+A(q_{\|}^2+rq_{z}^2)-{\rm i}C \omega,
  \label{eq:1}
\end{eqnarray}
where $Q$ is the AF ordering vector, $A$ and $C$ are constant, $q_{\|}$ is the wave vector component parallel to the 2D plane, and $q_z$ is that perpendicular to the plane. $r$ is a parameter indicating the degree of 3D character. Introducing the mode-mode coupling among the SF modes with wave vectors around $Q$, the temperature dependence of $\chi(Q)$ is explicitly calculated. When the mode-mode coupling is strong, as may frequently be the case including 4$f$ heavy electron systems, we can approximately set a condition that the local amplitude of SF, including thermal and zero-point fluctuations, is constant \cite{rf:14}. Here we use this approximation which simplifies the following calculations significantly. This formalism for quasi-2D systems was developed earlier in connection with the problem of non-FL properties in the heavy electron system CeCu$_{6-x}$Au$_x$ \cite{rf:15}.

Now for convenience, we introduce the reduced inverse susceptibility $y$ and reduced temperature $t$, and replace the parameters $A$ and $C$ by $T_A$ and $T_0$, respectively, having the dimension of energy: 
\begin{eqnarray}
    &&y=1/2T_A \chi (Q), ~ t=T/T_0,
  \nonumber \\ 
    &&T_A=Aq_{\rm B}^2/2, ~ T_0=(A/2 \pi C) q_{\rm B}^2,
  \label{eq:2}
\end{eqnarray}
where $q_{\rm B} =2 \sqrt{\mathstrut \pi}  /a$ is the effective zone boundary vector and $a$ is the lattice constant in the plane. Thus we need to calculate $y$ for a given set of parameter values. For this purpose we need still another parameter indicating the distance from the QCP. That is 
\begin{eqnarray}
    y_0=y(t=0)
  \label{eq:3}
\end{eqnarray}
on the paramagnetic side of the QCP and 
\begin{eqnarray}
    y_0 &=& -(T_A/3T_0) (M/2)^2, \mbox{ for $r=0$},
  \label{eq:4} 
 \\
    y_0 &=& S_{\rm LN}^2 (y=0, T=T_{\rm N}), \mbox{ for $r \ne 0$},
  \label{eq:5}
\end{eqnarray}
on the AF side of the QCP, where $M$ is the ordered moment per magnetic atom in $\mu_{\rm B}$ and $S_{\rm LT}^2$ is the local amplitude of the thermal SF. We refer to ref. \onlinecite{rf:15} for explicit equations for $y$ and $t_{\rm N}$. 

In order to evaluate $T_{\rm c}$ we first calculate the normal Green's function with the self-energy due to the above-discussed SF. The dynamical susceptibility may be regarded as a parameterized form of the result of a microscopic SCR calculation by using a given band structure. We take a tight-binding model for a simple tetragonal lattice with the nearest and second-nearest neighbor transfers, $-t_1$ and $t_2$, respectively, in the basal plane and a small transfer $t_z$ along the tetragonal axis which we neglect in practice without meaningful influence on the final results. 

The strong coupling calculation of $T_{\rm c}$ is performed in a standard way by the same procedure as that used before for the 2D and 3D problems \cite{rf:5,rf:6}. In numerical works we need a few more parameter values, i.e., approximate bandwidth: $W=8t_1$, onsite interaction in units of $W$:  $u=U/W$, $W/T_0$, $T_A/T_0$, and $t_2/t_1$. Following the previous estimations we here take the following values as a typical example: $u=0.7$, $W/T_0=5$, $T_A/T_0=4$, and $t_2/t_1=0.35$. 

\begin{figure*}[t]
  \begin{center}
    \includegraphics[scale=1.2]{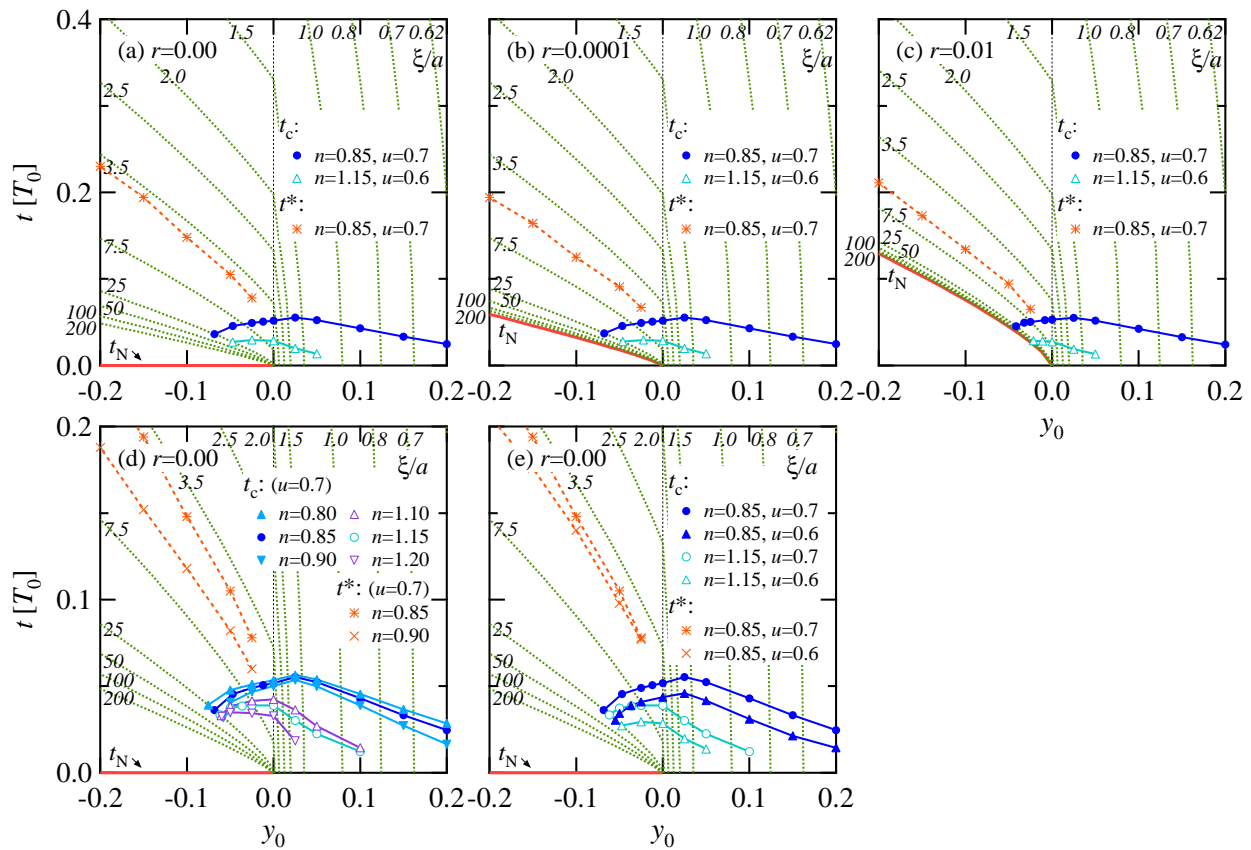}
  \end{center}
  \caption{
     Phase diagrams around AFQCP ($y_0=0$) with contour maps for the AF correlation length $\xi/a$, $T_{\rm N}$, $T_{\rm c}$, and the pseudogap onset temperature $T^*$. $y_0$ indicates the distance from the AFQCP and is approximately proportional to the doping concentration with the origin at the AFQCP, $t=T/T_0$, and $r$ indicates the degree of 3D character (see text). The marked change in the pattern of $\xi/a$ contours as $y_0$ changes sign is characteristic to quasi-2D AF metals. 
  }
  \label{fig:1}
\end{figure*}

\begin{figure}[t]
  \begin{center}
    \includegraphics[scale=1.2]{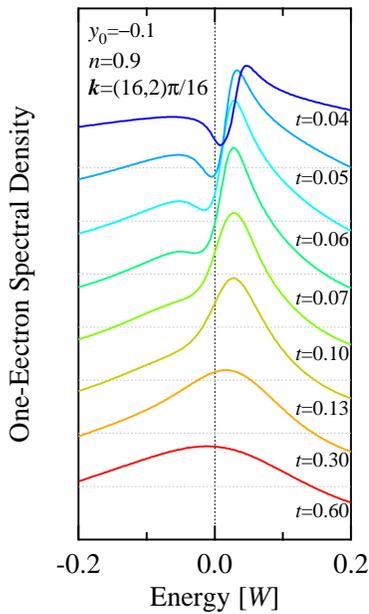}
  \end{center}
  \caption{
     One-electron spectral density for varying temperature at ${\bf k}=(\pi,2\pi/16)$, $n=0.9$, and $y_0=-0.1$. Pseudogap behavior is seen. 
  }
  \label{fig:2}
\end{figure}

\begin{figure}[t]
  \begin{center}
    \includegraphics[scale=1.2]{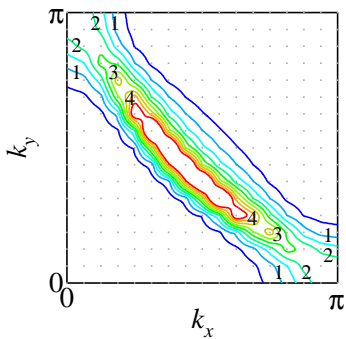}
  \end{center}
  \caption{
     Contour map for the one-electron spectral density at the Fermi energy, $n=0.9$, $y_0=-0.1$, and $t=0.625$, which is reminiscent of the Fermi arc.
  }
  \label{fig:3}
\end{figure}

We now present the results of the calculation. We first show the phase diagrams in the $t$-$y_0$ plane in place of those in the $T$-$x$ plane for doped cuprates, where $x$ is the doping concentration. $y_0$ just around the AFQCP is considered to be proportional to $(x-x_{\rm c})$, $x_{\rm c}$ being the critical concentration for the AFQCP. For larger $\left| y_0 \right|$ we only see the trend of concentration dependence. For $y_0<0$ the corresponding value of the AF moment $M$ may be evaluated approximately for small $r$ from eq. (\ref{eq:4}). We show in Fig. \ref{fig:1} $T_{\rm N}$, $T_{\rm c}$, and a contour map for the AF correlation length $\xi$:
\begin{eqnarray}
    \xi/a = (4 \pi y)^{-1/2},
  \label{eq:6}
\end{eqnarray}
where $a$ is the lattice constant of the 2D square lattice of magnetic atoms. Figures. \ref{fig:1}(a)-\ref{fig:1}(c) show the phase diagrams for $r=0$ (purely 2D system), $0.0001$, and $0.01$, respectively. For these small values of $r$ the contour maps are nearly the same except for lines in the vicinity of $T_{\rm N}$. $T_{\rm c}$ is calculated for the electron occupations corresponding to the e-doped ($n>1$) and h-doped ($n<1$) systems. Since $T_{\rm c}$ depends very weakly on $r$ we show in Figs. \ref{fig:1}(d) and \ref{fig:1}(e) more detailed results for $n$- and $u$-dependences when $r=0$. Although in reality the value of $n$ changes with $y_0$, we rather show here the weak $n$-dependence of the calculated $T_{\rm c}$ in Fig. \ref{fig:1}(d). In Fig. \ref{fig:2} we show the temperature dependence of the one-electron spectral density at the $(\pi,0)$-$(\pi,\pi)$ Fermi line crossing for $y_0=-0.1$ and $n=0.9$, where the temperature dependence of the chemical potential is neglected. Note that the $\bf k$-integrated spectral density also shows a similar pseudogap. The values of $T^*$ in Fig. \ref{fig:1} are estimated from the same type of drawings as Fig. \ref{fig:2} with more detailed $t$-dependence and for varying values of $y_0$ (not shown). The pseudogap onset is a gradual crossover and $T^*$ is an approximate measure for the crossover region, estimated here from the onset of an inflection point in the relevant part of the spectral function. Figure \ref{fig:3} shows a contour map of spectral density at the Fermi energy, which is reminiscent of the Fermi arc. Calculations for Figs. \ref{fig:2} and \ref{fig:3} are performed for $r=0$. 

We now discuss these phase diagrams in comparison with experimental results on high-$T_{\rm c}$ cuprates. We will not try to pursue better fit with experiment by parameter tuning, although it maybe possible. 

(1) The values of $T_{\rm c}$ in the h-doped regime are higher than those in the e-doped regime. This may be understood from the positions of the cross points (hot points) between the Fermi lines and the AF zone boundary line. The hot point for the former is closer to $(\pi,0)$, the van Hove singular point, where the DOS is large. In actuality the SF parameters may be somewhat different depending on the substance and the results here show the general trend and the order of magnitude. 

(2) The $T_{\rm c}$-$y_0$ curves are dome shaped with maxima around the AFQCP ($y_0=0$). The reduction of $T_{\rm c}$ with increasing $\left| y_0 \right|$ on the side of $y_0<0$ may be due to the formation of the pseudogap, which reduces the electron density at the Fermi level, and the increase in low frequency components of AFSF which enhances the effect of depairing and also reduces the intensity of high frequency components as the main contributor to the pairing. 

(3) The experimentally obtained phase diagram for e-doped cuprates in Fig. \ref{fig:1} of ref. \onlinecite{rf:8} is similar to Fig. \ref{fig:1}(c) ($r=0.01$), obtained from the present calculation with the contour map for the AF correlation length, the value for $T_{\rm N}$ and the $T_{\rm c}$ values for $n=1.15$ and $u=0.6$. 

(4) For the h-doped cuprates, if we assume that the optimal concentration is located at the AFQCP the relative positions of $T_{\rm c}$ and $T^*$ seem to be roughly consistent with those obtained experimentally, although the values for $T^*$ seem to be somewhat distributed depending on the experiment. A key question here is if the hidden AFQCP is really located at around the maximum of $T_{\rm c}$. According to the traditional phase diagram, the AF phase upon doping does not touch the SC phase and a spin-glass-like phase appears between them. It may be possible to conjecture that this spin-glass-like phase may intrinsically be a spin density wave (SDW) phase of the wavevector $q$ slightly different from the $Q$ of AF and that it persists up to the optimal concentration region when the SC phase is absent. The relatively low $T_{\rm N}$ of this possible SDW phase might be due to its smaller moment and/or weaker 3D character than that in the AF phase. Several recent experimental results seem to be in favor of the above conjectures for the position of the AFQCP \cite{rf:16,rf:17,rf:18,rf:19,rf:20,rf:21,rf:22-1,rf:22-2}.

One such result is the electrical resistivity of NdLSCO ($T_{\rm c}<20~{\rm K}$) under a very strong magnetic field \cite{rf:16}. $T$-linear behavior was observed down to $T = 0$ only for a hole concentration of $p^*=0.24$.  Although $p^*$ was interpreted as a pseudogap QCP, it may also possibly be interpreted as an AFQCP. Other results are the observation of quantum oscillations under extremely strong magnetic fields in the electrical and Hall resistance of certain cuprates \cite{rf:17,rf:18,rf:19,rf:20,rf:21}. The results indicate the existence of a small electron Fermi pocket and its possible interpretation may be the occurrence of an AF order. Another relevant experiment is nuclear magnetic resonance (NMR) performed on some cuprates with multi-layers of CuO$_2$ \cite{rf:22-1,rf:22-2}. Antiferromagnetic phases and coexistent AF-SC phases were observed in some of the layers with a hole concentration of up to $p=0.15$.

Finally, we note that we have not discussed the ordered phases below $T_{\rm N}$ and $T_{\rm c}$. Since the $T_{\rm N}$ and $T_{\rm c}$ curves cross at a certain point, the phase transition between the two phases may be of the first order or through a coexistent phase. The situation depends on the band structure of each substance and a fully microscopic theory is required to deal with this problem. 

In conclusion we have studied magnetism and superconductivity in quasi-2D metals around the antiferromagnetic quantum critical point. The obtained phase diagrams of $T_{\rm N}$, $T_{\rm c}$, and pseudogap onset temperature $T^*$ and/or contour maps for the AF correlation  length seem to be consistent with the experimental results for e-doped cuprates and also for h-doped cuprates, provided that the hidden AFQCP is located around the optimal doping concentration. The calculated $T_{\rm c}$-curves are dome shaped as generally observed in h-doped cuprates. Below $T^*$ a pseudogap in the one-electron spectral density is found in the high DOS region around $(\pi,0)$ and the contour map of the one-electron spectral density at the Fermi energy is reminiscent of the Fermi arc. In view of the simple and familiar model and approximations used here, the results seem to be significant indicating the importance of the spin fluctuations in quasi-2D metals around the AFQCP as one of the key concepts of the high-$T_{\rm c}$ problem. In particular, pseudogaps seem to be associated with quasi-2D AF metals. These concepts may be important in a wider range of superconductors including the recently discovered series of materials with FeAs layers \cite{rf:23}.

Provided that the above conjecture about the AFQCP in h-doped cuprates is valid, many major issues in the high-$T_{\rm c}$ problem may be understood. However, there are several outstanding experimental results still to be explained considering further details and/or some additional mechanisms. For example, SC fluctuations in addition to SF were considered to explain the pseudogaps observed in some of the transport phenomena including the Nernst anomaly and in NMR $T_1$ \cite{rf:24,rf:25,rf:26}. Further experimental and theoretical investigations are desired.  

Although the present approach using phenomenological expressions for the dynamical susceptibility is useful for obtaining overall pictures of the subject, it remains as an important future problem to develop a fully microscopic theory calculating dynamical susceptibilities from a given band structure. The fully microscopic SF theory, as extended to deal with the AF phases, may be used for this purpose \cite{rf:27}. On approaching the AF insulator phase we need to further extend the approach to cover the situations where the ground state has a relatively large ordered moment. This may belong to the main issue in the general theory of itinerant electron magnetism \cite{rf:28}.

We would like to thank K. Ueda for stimulating discussions and comments on the manuscript. 


\end{document}